# Photonic-chip-based generation of sub-100-femtosecond optical frequency combs

**Weiqiang Xie[1, 4, *], Zhengshun Lei[1, 4], Zeyu Xiao[2], Yudi Zhao[1], Xing Zou[2], Wenqi Wei[3], Zihao Wang[3], Ting Wang[3], Jianjun Zhang[3], Bofang Zheng[2,*], and Yikai Su[1]**

*1. State Key Laboratory of Photonics and Communications, School of Information and Electronic Engineering, Shanghai Jiao Tong University, Shanghai 200240, China*
*2. 2012 Labs, Huawei Technologies Co., Ltd., Shenzhen 518129, China*
*3. Beijing National Laboratory for Condensed Matter Physics, Institute of Physics, Chinese Academy of Sciences, Beijing 100190, China*
*4. These authors contributed equally to this work*
**weiqiang.xie@sjtu.edu.cn*

## Abstract

Sub-hundred-femtosecond optical pulses and frequency comb sources have been revolutionizing a wide range of applications, from ultrafast optical science to optical frequency standard and measurement. To date, the leading techniques for generating such pulses in practical systems rely on tabletop mode-locked lasers, which inherently suffer from high system complexity, limited long-term reliability, and pronounced environmental sensitivity. Meanwhile, driven by advances in photonic integration, chip-scale approaches have sought to realize miniaturized pulse sources. However, simultaneously achieving sub-hundred-femtosecond duration, ideal pulse shape, and a broadband flat-topped spectrum remains a significant challenge. Here, we address these challenges by combining two key photonic chip technologies: integrated thin-film lithium niobate electro-optic (EO) modulators for picosecond seed pulse generation[1,2], and highly nonlinear optical loop mirrors (NOLM)[3] based on aluminum gallium arsenide-on-insulator (AlGaAsOI) nanowaveguides for efficient temporal pulse cleaning and spectral broadening[4,5]. In theoretical simulation and experiment, we show that for an input EO picosecond seed pulse centred at approximately 1550 nm, a single-stage AlGaAs NOLM with a loop length of only one centimeter can produce flat-topped, nearly tenfold spectral broadening and over tenfold compression of the Fourier transform-limited pulse width, and more than 10 dB suppression of pulse pedestals. Using initial EO comb pulses with picosecond-level durations at repetition rates of 10–20 GHz, we demonstrate photonic-chip-enabled, low-noise pulses with an unprecedented duration of 55 fs and a flat-topped comb spectrum whose 10-dB optical bandwidth exceeds 90 nm. Our results highlight the remarkable potential of photonic chip technologies to realize high-repetition-rate, miniaturized sub-hundred-femtosecond optical pulse generators with the prospect of superior stability and operability. The demonstrated photonic-chip-based sub-hundred-femtosecond optical frequency comb sources may establish a new paradigm for both scientific research and practical applications.

## Main

Femtosecond mode-locked pulses and optical frequency comb sources, particularly those operating in the sub-hundred-femtosecond regime, deliver ultrashort, high-peak-power optical pulses with broad optical spectra. They have become indispensable tools across a wide range of disciplines and applications, including ultrafast spectroscopy[6] and microscopy[7], optical frequency metrology[8] and optical atomic clocks[9,10], microwave photonics[11-13], optical communication[14,15], ranging and distance metrology[16,17], and astronomy[18]. To date, the generation of sub-hundred-femtosecond optical pulses has primarily relied on tabletop solid-state crystal or fiber mode-locked lasers, such as the well-established Kerr-lens mode-locked Ti:sapphire lasers[19] and Er:fiber mode-locked lasers[20,21], where mode-locking occurs within a laser cavity. Beyond these cavity-based approaches, optical pulses can also be synthesized via non-resonant electro-optic (EO) modulation of a continuous-wave (CW) laser source, producing so-called EO combs[22-25]. While EO comb pulses typically fall within the picosecond regime, additional fiber-based nonlinear spectral broadening and pulse compression can further reduce their durations to the sub-hundred-femtosecond level[26,27]. Despite their considerable success, existing tabletop techniques for sub-hundred-femtosecond pulse generation still suffer from pronounced limitations in system complexity, miniaturization, and environmental sensitivity, thereby hindering reductions in size, weight, power and cost (SWaP-C) as well as improvements in stability, operability and long-term reliability.

To overcome these limitations, the rapid advancement of photonic integration technology over the past decades has opened a promising route toward chip-based optical pulse generators, spurring the development of diverse photonic-chip-based mode-locked sources. The first approach employs electrically pumped semiconductor mode-locked lasers that are monolithically or heterogeneously integrated on photonic chips. However, their output pulse widths typically remain at the picosecond level[28-30]. Alternatively, optical pulses can be generated via microresonator frequency comb sources (microcombs), which have been successfully demonstrated on various photonic material platforms over the past decade[31,32]. Yet, their practical utility is constrained by low efficiency, inherently poor spectral flatness, and limited stability and tunability[33]. More recently, the rapid progress in thin-film lithium niobate (TFLN) and lithium tantalite (TFLT) photonic integration has reignited interest in integrated EO combs based on on-chip modulators, yielding impressive pulse performance with pulse durations of a few hundred femtoseconds and flat-topped comb spectra[1]. Nevertheless, further reducing the pulse width into the sub-hundred-femtosecond regime remains a significant challenge. Consequently, current chip-scale optical pulse generators still face substantial hurdles in simultaneously achieving sub-hundred-femtosecond duration, ideal pulse shapes with suppressed pedestals, and broadband flat-topped comb spectra, impeding the advancement of chip-scale femtosecond optical frequency combs in both research and technology.

In this work, we address these challenges by leveraging two key photonic chip technologies: integrated TFLN EO modulators for the generation of GHz-repetition-rate, picosecond-level seed pulses, and low-loss, highly nonlinear aluminum gallium arsenide-on-insulator (AlGaAsOI) nanowaveguides for Kerr-based nonlinear spectral broadening. Crucially, we introduce the nonlinear optical loop mirror (NOLM) concept into the AlGaAsOI waveguides, enabling simultaneous spectral broadening and temporal pulse shaping. With an appropriate choice of NOLM parameters including splitting ratio, waveguide dispersion, and loop length, theoretical simulations indicate that for input EO picosecond seed pulses with picojoule-level energies, a single-stage AlGaAs NOLM with a loop length of only one centimeter can achieve flat-topped, nearly tenfold spectral broadening and high-fidelity, over tenfold compression of the Fourier transform-limited pulse duration. Moreover, the transmitted pulse after the AlGaAs NOLM exhibits over 10 dB suppression of pedestals compared with the initial pulse. Experimentally, for an input EO comb pulse with an initial duration of 1.17 ps at a 10 GHz repetition rate and a 10-dB optical bandwidth of 6 nm centred at approximately 1550 nm, the output pulse achieves a duration of 106 fs and a broadened optical bandwidth of 45 nm, together with a marked pedestal suppression. When the input pulse duration is reduced to 0.72 ps, the corresponding output pulse duration reaches 69 fs. For a seed pulse at a 20 GHz repetition rate, we further demonstrate a photonic-chip-enabled, low-noise pulse with an unprecedented duration of 55 fs and a flat-topped comb spectrum exceeding 90 nm. The key photonic component of AlGaAs NOLM occupies a footprint of only $200 \times 730\ \mu m^2$. Consequently, a fully integrated, chip-scale, sub-hundred-femtosecond pulse generator can be envisioned through incorporation with other photonic functionalities such as thin-film lithium niobate modulators, III-V lasers and on-chip amplifiers, thereby substantially advancing the development of chip-scale femtosecond optical frequency combs for both research and applications.

## Principle of femtosecond pulse generation based on TFLN and AlGaAs chips

Fig. 1a illustrates the principle of hundred-femtosecond pulse generation. A CW single-frequency laser is first injected into a TFLN photonic chip, where an initial EO frequency comb is generated via an on-chip time-lens system comprising one intensity modulator and three phase modulators driven by a single radio-frequency (RF) source. The resulting EO comb pulse exhibits a repetition rate of 10–20 GHz, determined by the driving RF frequency $f_m$. After dispersion compensation fiber (DCF) and amplification in an erbium-doped fiber amplifier (EDFA), the initial EO pulse is compressed to a nearly transform-limited duration of 0.5–1.2 picoseconds, with pulse energies of 5–10 picojoules, as depicted in Fig. 1a. This seed pulse is then coupled into the AlGaAsOI NOLM chip via a fiber-to-edge coupling scheme. A directional coupler (DC) splits the input pulse into two beams, which undergo spectral broadening through self-phase modulation (SPM) during propagation in the waveguide loop. Importantly, the NOLM exhibits a power-dependent nonlinear transmission response arising from the Kerr-induced nonlinear phase shift and the resulting interference, analogous to saturable absorption. This property effectively suppresses low-power pulse pedestals in the time domain. Finally, by applying dispersion compensation through a length-optimized single-mode fiber section, the output pulse can be further compressed to a nearly transform-limited duration on the order of hundreds or sub-hundred femtoseconds.

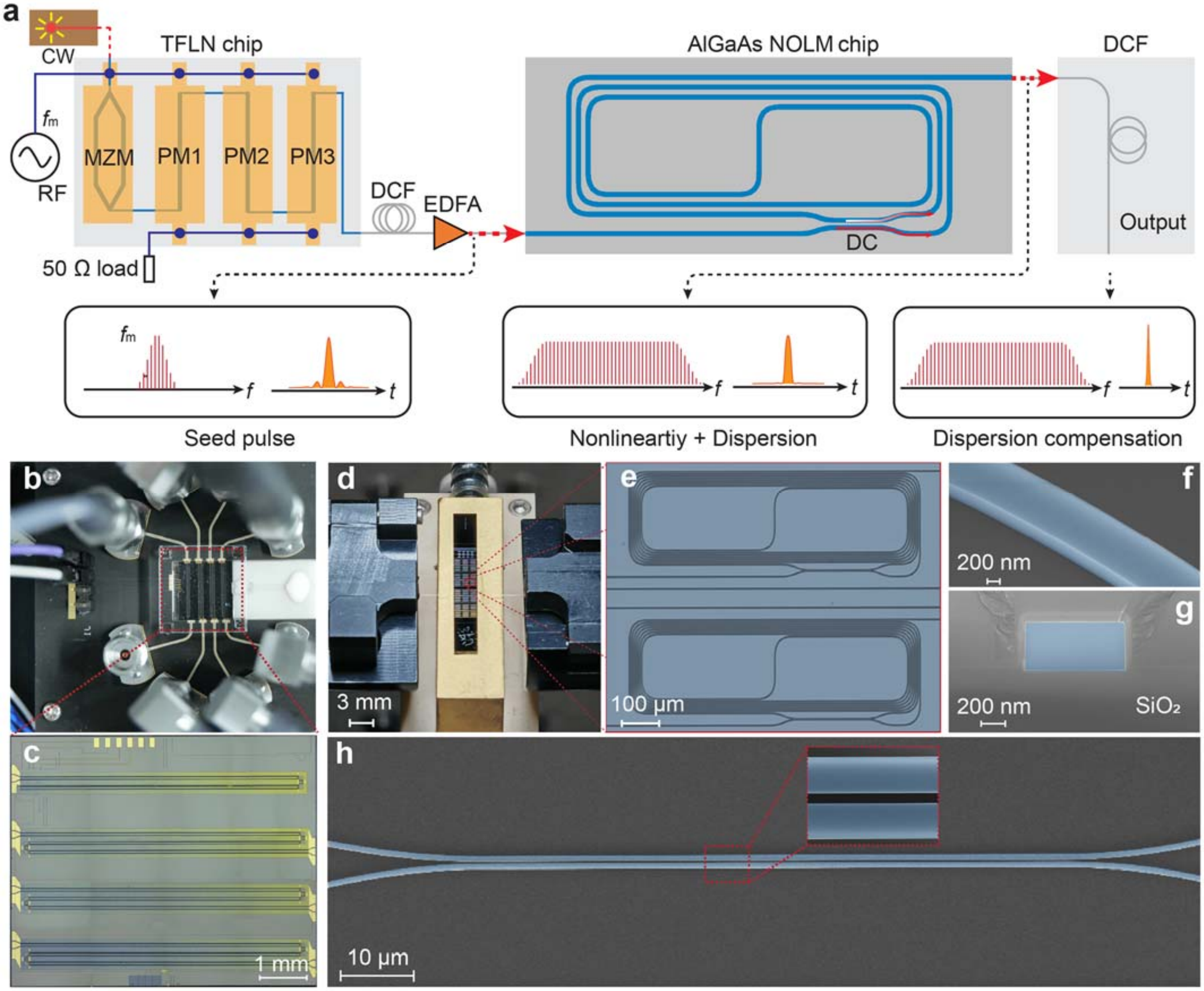

**Fig. 1 | Architecture of femtosecond pulse generation based on TFLN and AlGaAs photonic chips. a,** Schematic of the ultrashort femtosecond pulse generator. A CW laser is first modulated by a TFLN chip driven by an RF source to generate a coherent EO comb pulse. The initial EO comb pulse is then pre-compressed through DCF and amplified by an EDFA. The resulting EO seed pulse is coupled into an on-chip AlGaAs NOLM, where the SPM process enables effective spectral broadening and simultaneously imposes a nonlinear chirp on the pulse. Finally, the output pulse is compressed by a DCF to produce ultrashort pulse with durations in the hundred-femtosecond regime. **b, c,** Photograph of the fully packaged TFLN EO comb generator with

optical input/output fibers and electrical RF interfaces (**b**) and zoomed-in micrograph of the LN photonic chip consisting four modulators (**c**). **d, e,** Optical micrographs of the AlGaAs NOLM chip under test (**d**) and a magnified view of the AlGaAs NOLM devices (**e**). **f, g,** SEM images of an AlGaAs waveguide without the top $SiO_2$ cladding (**f**) and the waveguide cross section with the $SiO_2$ cladding (**g**). **h,** Top-view SEM image of an AlGaAs DC. Inset: zoomed-in view of the coupled waveguides separated by a gap.

In our experimental demonstration, the TFLN photonic chip was fabricated using optimized processes (Methods) and subsequently packaged compactly with input/output optical fibers and RF electrical interfaces, as illustrated in Fig. 1b and 1c. The TFLN EO comb generator produces short pulse trains with a high, tunable repetition rate and excellent spectral flatness[2]. However, owing to insufficient modulator efficiency and limited modulation depth even under high microwave driving powers, the spectral bandwidth of the initial EO comb remains narrow, typically resulting in picosecond or sub-picosecond pulse durations.

Inspired by conventional tabletop fiber-based NOLM systems for nonlinear spectral broadening and pulse shaping[3,34,35], we extend this concept, for the first time, to a photonic chip using nonlinear waveguides. To accumulate sufficient nonlinear phase shift in a practical NOLM, traditional integrated dielectric waveguides require loop lengths on the order of meters. Such extended lengths introduce considerable propagation loss and dispersion accumulation, which dissipate pulse peak power and distort the pulse waveform, ultimately impairing spectral broadening and pulse shaping. In contrast, emerging low-loss, high-confinement AlGaAsOI waveguides[4,5,36,37] offer an effective nonlinear coefficient two orders of magnitude higher, enabling centimeter-scale loop lengths driven by modest peak powers of a few watts. This key advantage, together with low propagation loss, effectively mitigates the detrimental impact of loss and dispersion on nonlinear broadening, establishing AlGaAsOI as an ideal platform for on-chip NOLMs in the telecom band.

Compared with fiber-based NOLMs, AlGaAs photonic NOLMs enable both the geometric miniaturization of fiber loops from hundreds of meters to on-chip centimeter-long waveguides, and the reduction of driving peak powers from the kilowatt to the watt level. More importantly, AlGaAs NOLMs overcome inherent physical limitations arising from long-distance pulse propagation. For instance, fiber-based NOLMs can introduce significant dispersion accumulation, polarization drift, and propagation delay, all of which pose considerable challenges for practical design and application[38,39]. Owing to the short waveguide length and flexible mode management, these limitations are largely eliminated in AlGaAs NOLMs. Finally, whereas the bulky implementation of long fiber loops renders the system highly susceptible to environmental influences and compromises long-term stability and reliability, the compactly integrated AlGaAs NOLMs inherently exhibit strong resistance to external disturbances and thus offer superior stability.

For the AlGaAsOI waveguide platform, we adopt a heterogeneous wafer-bonding technique to transfer the AlGaAs film epitaxially grown on a GaAs substrate, combined with an optimized low-loss waveguide fabrication process (Methods). The waveguide height is fixed at 350 nm, while the width is varied to tailor the group-velocity dispersion (GVD) of the fundamental transverse-electric (TE) mode from the anomalous to the normal regime. Based on quality-factor measurements of ring resonators and four-wave mixing experiments in straight waveguides, the propagation loss is estimated to be ~0.6 dB/cm, and the effective nonlinear coefficient ($\gamma$) is extracted as 437 $W^{-1}m^{-1}$. Given the input pulse conditions in our experiment, the key design parameters for optimal spectral broadening and pulse shaping in the AlGaAs NOLM are the waveguide width, the splitting ratio ($\rho$) of the DC (i.e., the power coupling ratio of cross port), and the physical loop length ($L$). To suppress modulation instability and minimize dispersion-induced temporal broadening, the entire NOLM structure is engineered to operate in the weak normal-dispersion regime, with typical waveguide widths of 750–800 nm. To determine the remaining parameters, we first analyze the static power-dependent transmission function, and then employ the generalized nonlinear Schrödinger equation (GNLSE) to numerically simulate the dynamic pulse evolution in both time and frequency domains. Accordingly, the splitting ratio is set to $\rho \approx 0.35$–$0.45$, and the loop length to $L \approx 0.7$–$3.0$ cm. Notably, Euler bends are employed in the loop to maintain a compact footprint while ensuring a smooth, low-loss transition for the fundamental TE mode.

Our simulations demonstrate that for picosecond seed pulses with picojoule-level energies (peak powers below ten watts), a single-stage AlGaAs NOLM can achieve flat-topped, nearly tenfold spectral broadening, accompanied by high-fidelity, over tenfold compression of the transform-limited pulse width and temporal-pedestal suppression exceeding 10 dB. In the experimental demonstration, to accommodate potential fabrication deviations from the design, we fabricated a set of AlGaAs NOLM devices with varying waveguide widths, splitting ratios, and loop lengths to cover the estimated optimal parameter range for spectral broadening. Fig. 1d and 1e present optical photographs of the fabricated AlGaAs photonic chip and the NOLM device, which occupies a footprint of only 200 × 730 $\mu m^2$. The corresponding scanning electron microscope (SEM) images of the waveguide and its cross section are shown in Fig. 1f and 1g, while the top-view SEM image of the directional coupler is provided in Fig. 1h. The fabricated AlGaAs waveguides exhibit remarkably vertical and smooth sidewalls, ensuring both low loss and precise dispersion engineering consistent with the simulations.

## Efficient spectral broadening and temporal pulse shaping with AlGaAs NOLM

Fig. 2a presents the experimental setup for spectral broadening of an input picosecond-level EO comb pulse using the AlGaAsOI NOLM (Methods). The initial EO comb pulses, generated by the TFLN chip, exhibit spectral bandwidths of 6–12 nm and tunable repetition rates of 10–20 GHz. After power amplification and dispersion compensation, the seed pulses attain durations of 1.2–0.5 ps and average powers of up to 30 dBm. For various NOLM devices, we investigate the output pulse characteristics by varying the input pulse power to determine the optimal operating conditions for spectral broadening. Comprehensive characterization reveals that optimal spectral broadening, combining a large spectral bandwidth with high flatness, is obtained with waveguide widths of ~750–800 nm, a splitting ratio of $\rho \geq 0.4$, and a loop length of ~1.0 cm, in good agreement with our simulations. Based on these results, we select appropriate AlGaAs NOLM devices and subsequently characterize the output pulses in both the frequency and time domains.

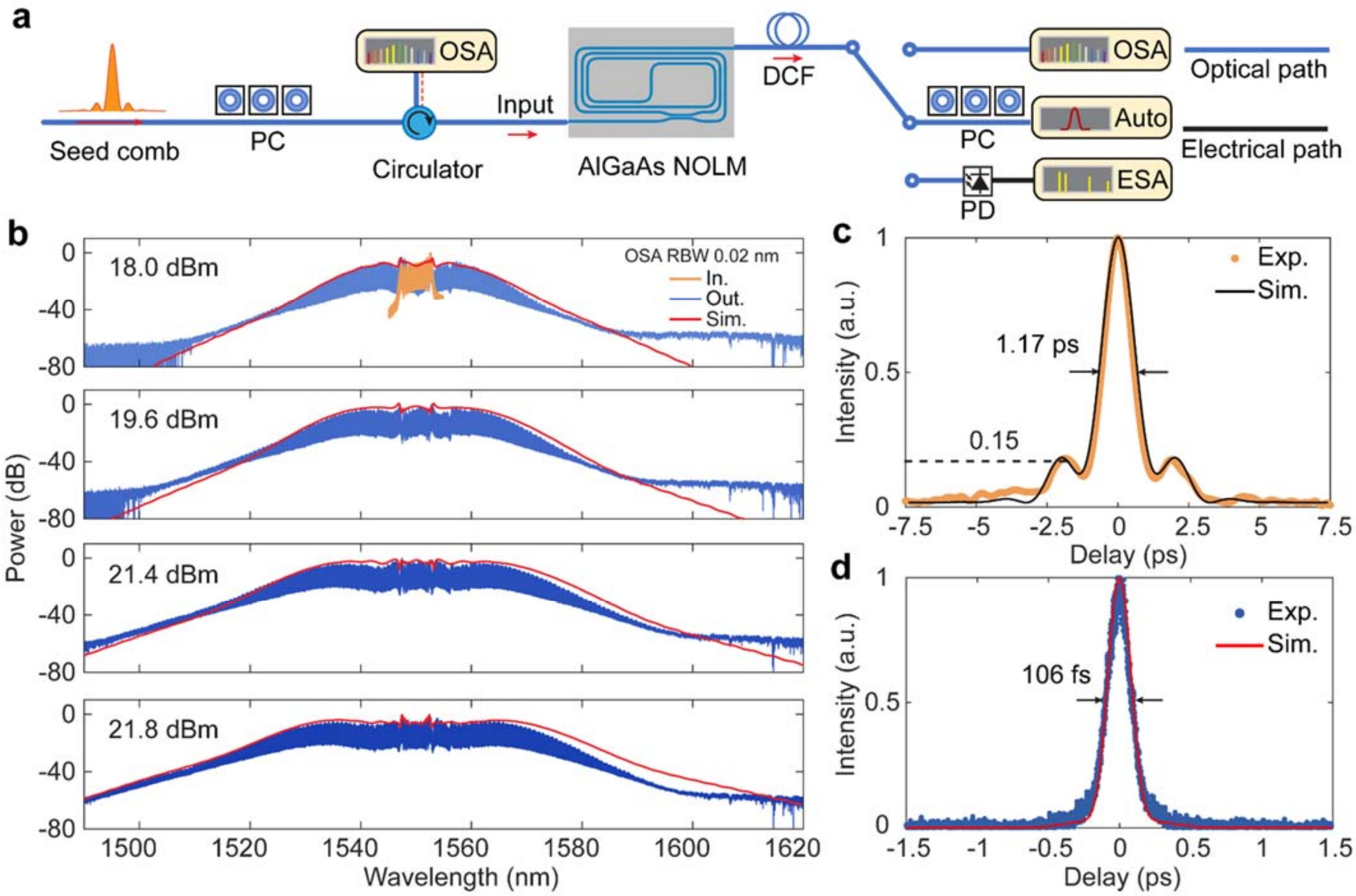

**Fig. 2 | Experimental setup and measurement results of pulse spectral broadening and temporal compression. a,** Schematic of the experimental setup for pulse broadening and compression of pulse in the AlGaAs NOLM. PC, polarization controller; OSA, optical spectrum analyzer; Auto, autocorrelator; PD, photodetector; ESA, electrical spectrum analyzer. **b,** Measured comb spectra of the output pulses at different pump powers. The input seed EO comb has a bandwidth of ~6 nm centered at 1550 nm and a repetition rate of 10 GHz, as shown in the top panel (orange line). The corresponding simulation results are presented as red lines. **c,** Measured the temporal envelope of the seed pulse (orange dots) and the corresponding theoretical fit (black line). The pulse duration of 1.17 ps and the pedestal intensity of 15% are indicated. **d,** Measured temporal envelope of the output pulse (blue dots) at an optimal input power of 21.8 dBm, together with the theoretical fit (red line). The pulse duration of 106 fs is denoted.

First, we employ a configuration consisting of one intensity modulator and two TFLN phase modulators to generate seed EO comb pulses with a duration of 1.17 ps, a flat-topped spectral bandwidth of ~6 nm centered near 1550 nm, and a repetition rate of 10 GHz. The AlGaAs NOLM features a waveguide width of 750 nm and a splitting ratio of $\rho \approx 0.42$ at 1550 nm, determined either from a reference DC coupler or extracted from the linear transmission spectrum of the NOLM. We pump the NOLM at various seed pulse powers and examine the spectral evolution of the output pulse. For simplicity, the pulse power is characterized by the average power measured directly with an optical power meter. Fig. 2b presents the output comb spectra at on-chip pump powers of 18.0 dBm, 19.6 dBm, 21.4 dBm, and 21.8 dBm. The orange line in the first panel shows the seed comb spectrum for reference. Note that after amplification in the EDFA, the seed pulse exhibits a tilted spectral profile that degrades its flatness. At a pump power of 18 dBm, the comb spectrum broadens considerably, although intensity fluctuations remain pronounced, particularly at the wings of the seed comb. Increasing the pump power further enhances the broadening, yielding a 10-dB optical bandwidth of ~45 nm at 21.8 dBm, as shown in Fig. 2b. In addition, the spectral flatness improves dramatically, surpassing that of the seed comb. At pump powers above 21.8 dBm, the spectral bandwidth increases only slowly, while the flatness degrades considerably. This degradation occurs because the intensified SPM-induced chirp at elevated pump powers accelerates optical wave breaking in the normal dispersion regime[40]. Consequently, the optimal input pulse power for this NOLM device is determined to be 21.8 dBm.

We further perform numerical simulations using the GNLSE to reproduce the experimental results, as shown by the red curves in Fig. 2b. The simulated spectral profiles exhibit overall agreement with the experiment, indicating that the frequency-domain evolution in the AlGaAsOI waveguide-based NOLM is predominantly governed by Kerr nonlinearity and dispersion. At higher pump powers, a slight deviation between simulation and experiment is observed in the long-wavelength regime (>1570 nm), which is primarily attributed to wavelength- or power-dependent nonlinear losses[41]. Incorporating these nonlinear losses into the simulation model yields good agreement with the experiment, as detailed in the subsequent analyses.

Following the frequency-domain characterization, we examine the output pulse in the time domain to investigate the effect of the NOLM on the temporal envelope. For reference, the normalized seed pulse envelope is shown in Fig. 2c, with a measured duration of 1.17 ps and a pedestal level of approximately 15%, corresponding to a suppression ratio of 8.2 dB. The theoretical fit based on a time-lens system with cascaded EO modulators is also plotted as a solid black line. Such pedestals, with considerable power levels, are intrinsic to EO comb pulses. They not only reduce the energy fraction contained in the main peak but also prevent the pulse from reaching the Fourier transform limit. This can adversely affect subsequent pulse manipulation and applications, leading, for example, to dissipation of optical amplifier gain, degradation of spectral flatness, and increased optical noise. Since the output pulse carries a certain positive chirp after propagating through the AlGaAs NOLM, appropriate dispersion compensation can compress it to a nearly transform-limited duration. In the experiment, we employ a standard single-mode optical fiber with an optimized length of ~1.2 meters for this compensation. Fig. 2d presents the resulting output pulse profile at the optimal input power of 21.8 dBm, together with the corresponding theoretical fit. The measured pulse duration is 106 fs, close to the Fourier transform limit of 100 fs, demonstrating over tenfold compression relative to the input pulse width. More importantly, the pedestal level is suppressed to below 1% (limited by the measurement sensitivity of the autocorrelator), corresponding to an improvement of more than 10 dB in the pedestal suppression ratio.

## Sub-100-fs frequency comb generation

To further push the pulse duration below 100 fs, we adopt shorter, sub-picosecond input pulses and, assuming the same broadening factor demonstrated above, anticipate achieving output pulses well below one hundred femtoseconds. Theoretically, a shorter pump pulse also enables more efficient spectral broadening, as the SPM-induced chirp scales with the temporal intensity gradient of the pump pulse. To verify this, we employ a configuration of one intensity modulator and three TFLN phase modulators to generate seed EO comb pulses with narrower widths and larger spectral bandwidths. By optimizing the operating conditions of the driving RF signal and the modulators, we obtain seed pulses with sub-picosecond durations down to ~0.5 ps and comb bandwidths up to ~12 nm. Among the fabricated AlGaAs NOLMs with varying waveguide widths and splitting ratios, we select a device with a width of ~775 nm and a splitting ratio of $\rho \approx 0.45$ at 1550 nm, which delivers optimal performance in the following experiment.

First, we generate a seed EO comb pulse with a duration of 721 fs at a repetition rate of 10 GHz to pump the AlGaAs NOLM. The resulting output comb spectrum at an optimal on-chip pump power of 19.4 dBm is shown in Fig. 3a. The spectral bandwidth expands from an initial ~8 nm to over ~72 nm, yielding more than 900 comb lines within the 10-dB bandwidth, with the exception of a few residual lines at the wings of the seed comb. Furthermore, by driving the TFLN modulators with a 20-GHz RF signal, we generate a seed EO comb pulse with a duration of 488 fs and a spectral bandwidth of ~12 nm as the pump pulse for the AlGaAs NOLM. Fig. 3b presents the corresponding output spectrum at an optimal on-chip pump power of 22.7 dBm, demonstrating a broadened spectral bandwidth exceeding 90 nm. Note that both output spectra exhibit asymmetry with respect to the pump center, characterized by greater broadening in the short-wavelength regime than in the long-wavelength regime. This behavior is likely attributed to high-order nonlinear processes, particularly at elevated pump powers, such as three-photon absorption (3PA) and self-steepening (SS)[42,43], which break the symmetry of spectral broadening. To verify this, we incorporate these high-order nonlinear processes into our simulation model. The simulated spectral profiles, shown in Fig. 3a and 3b, agree well with the experimental results.

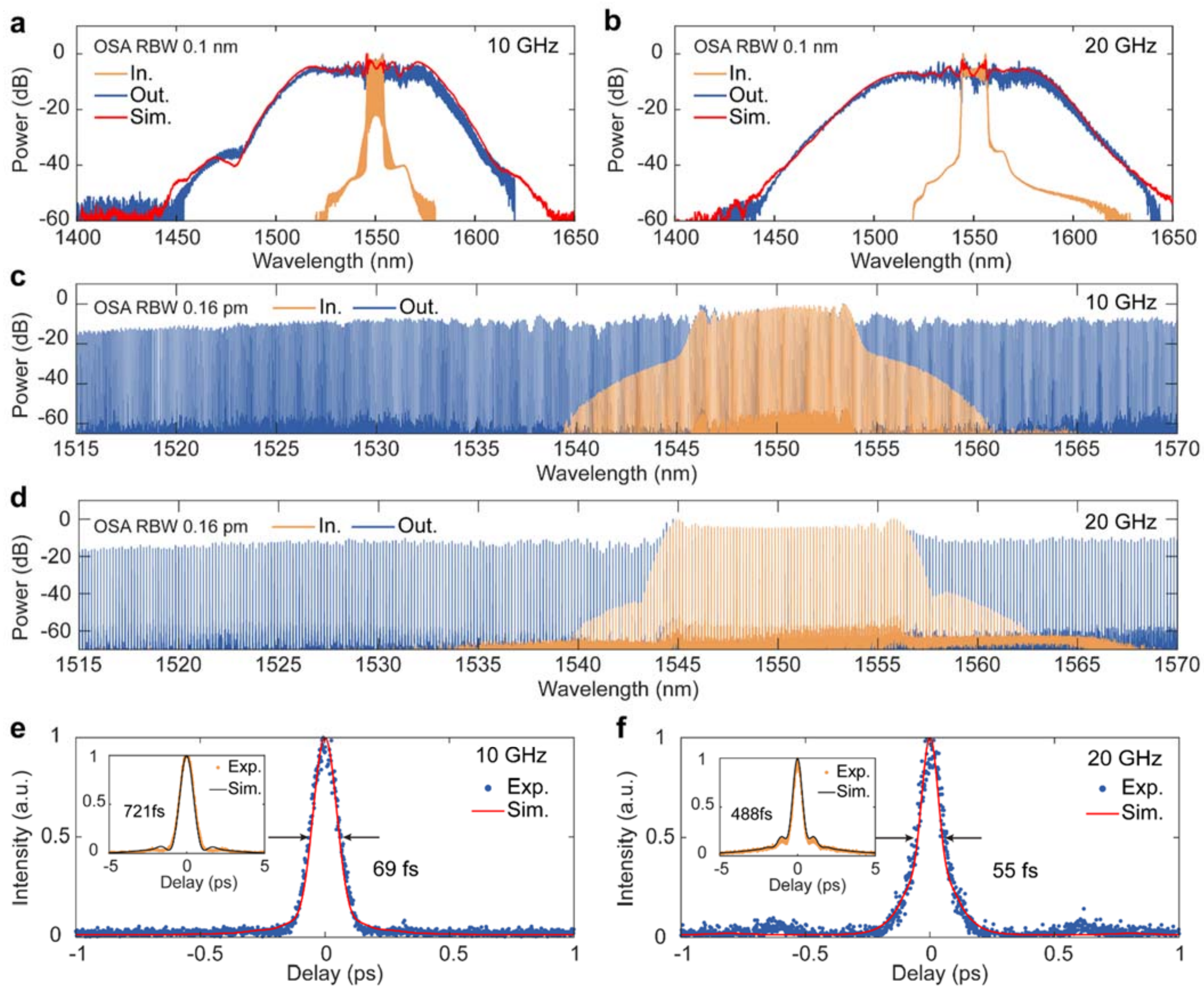

**Fig. 3 | Generation of sub-hundred-femtosecond frequency combs. a,b,** Measured output comb spectra (blue lines) for seed pulses at repetition rates of 10 GHz (**a**) and 20 GHz (**b**). The input seed pulse spectra (orange lines) and the corresponding simulation results (red lines) are also shown. **c,d,** High-resolution comb spectra of the input and output pulses presented in (**a**) and (**b**), respectively. **e,f,** Measured temporal envelopes (blue dots) and the corresponding theoretical fits (red lines) of the output pulses at repetition rates of 10 GHz (**e**) and 20 GHz (**f**). Insets show the measured and simulated results for the corresponding seed pulses. All pulse durations are indicated.

To further assess the optical quality of the comb lines, we employ a high-resolution optical spectrum analyzer to record the output spectra over a wavelength range of 1515–1570 nm, limited by our instrumentation, as displayed in Fig. 3c and 3d. Remarkably, the output comb spectra maintain exceptional flatness, except for a few lines at specific wavelengths. More importantly, the optical signal-to-noise ratio for nearly all comb lines exceeds 40 dB, a level comparable to that of the seed comb pulse, indicating that the coherence and optical noise characteristics of the comb lines are well preserved after spectral broadening in the NOLM.

Next, we characterize the temporal properties of the input and output pulses and show the results in in Fig. 3e and 3f for repetition rates of 10 GHz and 20 GHz, respectively. Prior to time-domain measurement, appropriate dispersion compensation using a standard single-mode optical fiber is applied to each pulse, aiming to achieve a nearly transform-limited duration. For the 721 fs input pulse at 10 GHz, the output pulse duration is reduced to 69 fs, close to the Fourier transform-limited value of 64 fs as shown in Fig. 3e, representing an over tenfold compression in pulse width. For the 488 fs input pulse at 20 GHz, the output pulse duration reaches 55 fs, as presented in Fig. 3f. The corresponding Fourier transform-limited duration is 42 fs, suggesting that further compression may be attainable through higher-order dispersion compensation. In addition, the output pulses not only retain a well-defined, distortion-free shape but also exhibit a significant reduction in pedestal level, from an initial 10–20% to 1–2% after the NOLM. To the best of our knowledge, this represents the first demonstration of photonic-chip-enabled sub-hundred-femtosecond optical pulses featuring flat-topped comb spectra, a high pedestal suppression ratio, and a high, tunable repetition rate.

Finally, to evaluate the impact of nonlinear broadening on the mode-locking performance of the output pulse, we measure the single-sideband (SSB) phase noise of the repetition-frequency beat signal using a high-speed photodetector and an electrical spectrum analyzer (ESA) (Methods). The results are presented in Fig. 4a and 4b for output pulses at repetition rates of 10 GHz and 20 GHz, respectively. For comparison, the corresponding results for the driving RF signal and the input pulses are also shown. The phase noise of the RF signal is measured directly by connecting the RF source to the ESA. Notably, the measured phase noise curves for both the input and output pulses nearly overlap with that of the original RF source at offset frequencies below ~$10^6$ Hz. This indicates that the mode-locking performance is well preserved during spectral broadening and temporal compression in the AlGaAs NOLM. At offset frequencies above $10^6$ Hz, the phase noise curves of the comb pulses trend upward relative to the RF source, which can be attributed to the residual noise floor introduced by processes such as electrical amplification of the RF signal, electro-optic modulation, and optoelectronic conversion in the photodetector. Additionally, the beat-note frequency spectra for the RF signal and the input/output pulses are shown in Fig. 4c and 4d for repetition rates of 10 GHz and 20 GHz, respectively. These spectra exhibit nearly identical signal-to-noise ratios, confirming that the excess noise added by the AlGaAs NOLM is negligible.

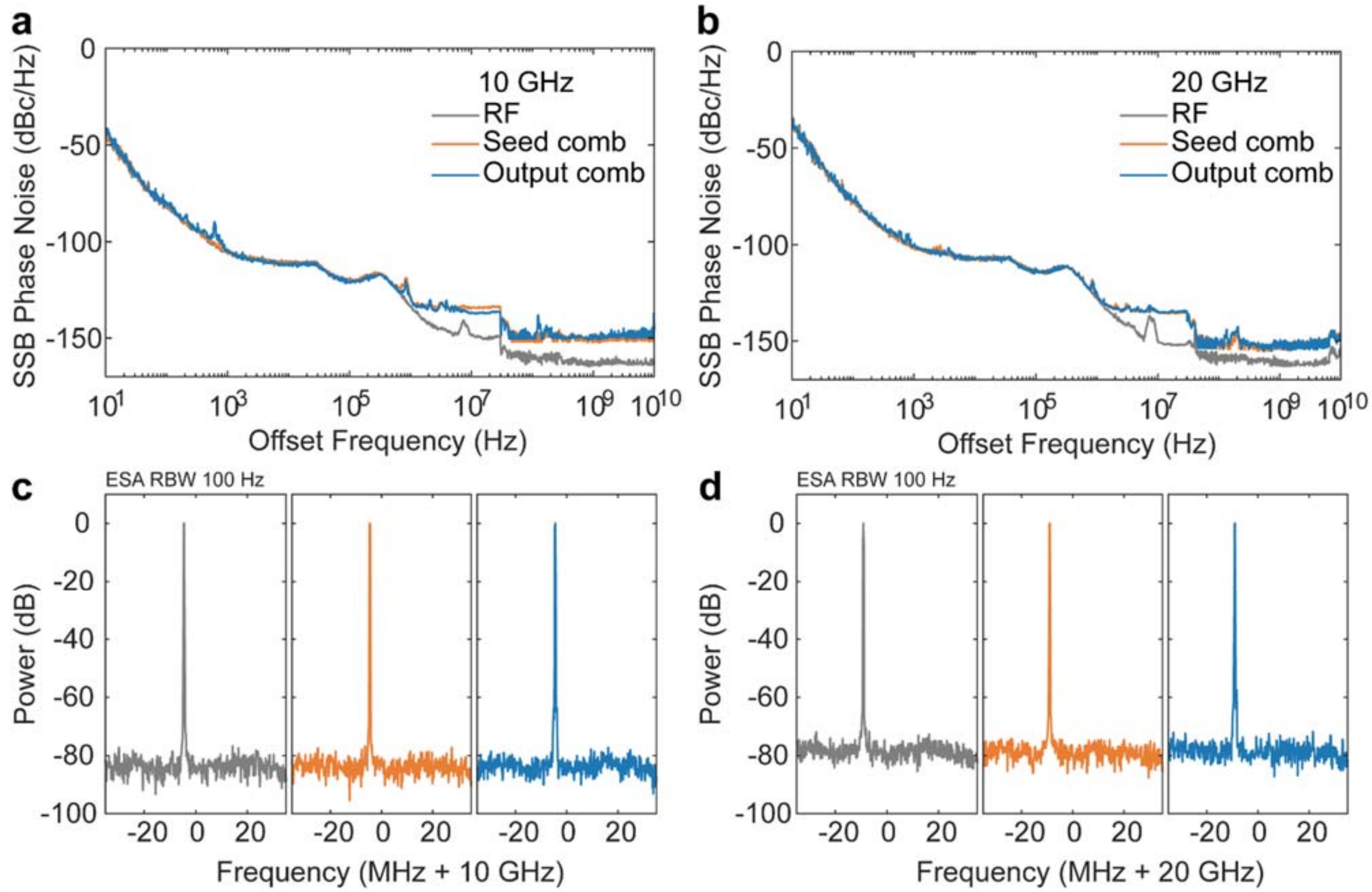

**Fig. 4 | Characterization of the phase noise for the repetition-frequency beat signal. a,b,** Measured SSB phase noise frequency spectra for the repetition-frequency beat signals of the output pulses at the repetition rates of 10 GHz (**a**) and 20 GHz (**b**). The results for the driving RF signal and the seed pulse are also shown for comparison. **c,d,** Measured beat-note spectra corresponding to the results in (**a**) and (**b**), respectively.

## Discussion and outlook

In this work, we have demonstrated chip-based, sub-hundred femtosecond pulse generation by combining the electro-optic effect of the TFLN photonics with the high third-order nonlinearity of AlGaAsOI photonics. The TFLN platform delivers picosecond-level seed EO comb pulses with excellent spectral flatness at tunable repetition rates ranging from several GHz to tens of GHz, while the highly nonlinear AlGaAsOI nanowaveguides enable ultra-efficient spectral broadening. In particular, we have introduced a NOLM mechanism into the AlGaAs waveguide design that simultaneously harnesses Kerr-induced spectral broadening and nonlinear interference, thereby providing a crucial degree of freedom for tailoring both spectral flatness and temporal pulse shape. Experimentally, using a single-stage AlGaAs NOLM with a loop length of only one centimeter (corresponding to a sub-square-millimeter footprint), we achieved flat-topped, nearly tenfold spectral broadening and over tenfold compression of the transform-limited pulse duration. This ultimately yields photonic-chip-enabled, low-noise, sub-hundred-femtosecond pulses with durations of 69 fs and 55 fs, and comb spectral bandwidths exceeding 70 nm and 90 nm, at repetition rates of 10 GHz and 20 GHz, respectively. Notably, owing to the nonlinear filtering effect of the NOLM, the output pulse pedestal is significantly suppressed, resulting in an over 10 dB improvement in the suppression ratio. Overall, these results establish a new benchmark for chip-based optical pulse generation, outperforming commercial tabletop pulse generators in multiple key aspects.

Compared with conventional fiber-based NOLMs, the demonstrated chip-based AlGaAs NOLMs reduce both the required driving power and loop length by several orders of magnitude, highlighting their exceptional nonlinear efficiency and advantage in compact integration. These integrated photonic NOLMs hold significant potential and broad application value. They not only enable on-chip, high-efficiency pulse manipulation in both the frequency and time domains, but also offer a new on-chip mode-locking strategy for universal optical pulse generation. Furthermore, the photonic-waveguide NOLM concept can be implemented on other photonic material platforms to further extend the operating wavelength bands, which will undoubtedly reshape the research and technological landscape of ultrashort optical pulses.

Despite its outstanding performance, the current NOLM device can be further improved in terms of reduced pump power, increased spectral bandwidth and flatness, enhanced operational tunability, and higher energy efficiency. These improvements can be achieved through optimization of several aspects, including AlGaAs waveguide loss and material composition, NOLM design (e.g., beam splitter and loop length), and fiber–chip coupling. Beyond the AlGaAs NOLM, other system components also offer room for optimization. First, the TFLN EO modulators are inherently capable of operating at higher speeds of up to 40 GHz and beyond, thereby enabling an increase in the pulse repetition rate. Additionally, the recently developed TFLT technology can be leveraged to extend the optical bandwidth of the EO seed pulse[44]. Second, the current dispersion compensation using optical fibers for input and output pulses can be replaced by on-chip dispersion management with photonic waveguides or gratings, further improving system compactness[1]. Finally, further reductions in NOLM waveguide loss and inter-component insertion loss will lower both the pump power requirement of the NOLM and the overall system power budget. Ultimately, we envision that a miniaturized, fully chip-scale, sub-hundred-femtosecond optical pulse generator may become feasible through integration with photonic light sources such as III–V lasers and on-chip amplifiers[45-48], as well as RF electrical driver chips[49,50].

In summary, our work demonstrates that the combination of the emerging photonic technologies of LN/LT-OI and AlGaAsOI can unify $\chi^{(2)}$-based EO and $\chi^{(3)}$-based Kerr effects, thereby establishing a coherent link between microwave signals and optical frequency combs with broad bandwidths and high coherence. This synergy offers an unprecedented capability for generating and manipulating ultrafast optical pulses in an all-chip manner. The demonstrated high-repetition-rate, sub-hundred-femtosecond optical frequency comb generator opens up new opportunities across diverse fields, ranging from ultrafast optics and spectroscopy to coherent optical datalinks and optical frequency metrology. Moreover, our work paves the way toward building highly compact, on-chip ultrashort femtosecond comb systems with the potential for superior stability, operability, long-term reliability, and reduced SWaP-C, thereby establishing a new paradigm for ultrafast optical science and the development of practical future technologies.

# Methods

**Device fabrication**

The AlGaAsOI waveguide NOLM devices are fabricated using a heterogeneous wafer-bonding technique mostly following our previously developed process[37]. A 350nm-thick $Al_{0.2}Ga_{0.8}As$ device layer on the top of 400nm-thick $Al_{0.8}Ga_{0.2}As$ etch-stop layer is epitaxially grown on a GaAs substrate. Then the AlGaAs device layer is covered with ~8nm-thick $Al_2O_3$ film by atomic layer deposition (ALD) and subsequently bonded onto a Si substrate with a 3μm-thick thermal $SiO_2$ layer. After removal of the GaAs substrate and etch-stop layer, the resulting AlGaAsOI film structure is sequentially deposited with ~8nm-thick $Al_2O_3$ and ~50nm-thick $SiO_2$ by ALD as a dry-etch hard mask for AlGaAs waveguide. The devices are patterned using electron-beam lithography, and the $SiO_2$ hard mask and AlGaAs are then etched in inductively coupled plasma reactive-ion etching (ICP-RIE) to define the AlGaAsOI waveguide NOLM structures. Finally, the devices are passivated with a 15nm-thick $Al_2O_3$ layer by ALD and clad with a 2μm-thick $SiO_2$ layer via plasma-enhanced chemical vapor deposition (PECVD). To ensure reliable fiber–chip coupling, the chip edges are mechanically diced and polished using chemical-mechanical polishing, yielding a coupling loss of approximately 3.0 dB per facet.

The integrated EO comb generator is fabricated on a commercial *x*-cut LNOI wafer (NANOLN) consisting of a 360-nm-thick LN film on a Si substrate with a 9μm-thick thermal $SiO_2$ layer. The LN waveguide passive components are defined using deep-ultraviolet (DUV) lithography, followed by dry etching of 180 nm LN via ICP-RIE. A 1.2-μm-thick $SiO_2$ top cladding is subsequently deposited by PECVD. The modulation region is then opened by DUV lithography, and 200 nm of the $SiO_2$ cladding is removed by ICP-RIE etching. A 100-nm-thick Ni/Cr layer is deposited and patterned via a lift-off process to form the thermo-optic phase shifters and on-chip resistors. Subsequently, a 900-nm-thick gold layer is deposited and patterned, also via lift-off, for the high-speed RF electrodes. Finally, the chip is cleaved and polished for edge coupling. After testing, the chip is packed with input/output optical fibers and RF connectors for system-level application. Further details of the device design and fabrication can be found in our previous work[2].

**Experimental setup for sub-femtosecond pulse generation**

The initial EO comb pulse is generated on the TFLN photonic chip using a single-frequency CW laser input, then amplified by an EDFA, and subsequently compressed to a picosecond-level seed pulse through dispersion compensation in a standard single-mode fiber (SMF) with a length of 100m-325m, depending on the EO comb configuration. The output power of the CW laser is set to 13 dBm. The TFLN chip has an insertion loss of approximately 11 dB, including fiber–chip coupling losses. After amplification by the EDFA, the average power ranges from ~20 to 26 dBm. The resulting seed pulse passes through a polarization controller (~2 m) and a fiber circulator (~2 m) to adjust polarization and monitor the reflected spectrum, respectively. It is then coupled into the AlGaAs NOLM chip via horizontal coupling between a lensed fiber (~1 m) and an inverse AlGaAsOI waveguide taper, with a coupling loss of approximately 3 dB per facet. The AlGaAs chip is mounted on a thermally controlled stage maintained at 20 ℃ to minimize temperature fluctuations and ensure measurement repeatability. After the NOLM, the output pulse is coupled into another lensed fiber (~1 m) and then transmitted through an SMF with an adjustable length of 0.5–3.0 m for dispersion compensation. The full comb spectra are recorded using an optical spectrum analyzer (Yokogawa AQ6370E), and the optical signal-to-noise ratio is determined with a high-revolution spectrum analyzer (APEX-AP2040A). The temporal pulse traces are measured with an autocorrelator (FR-103XL). For the single-sideband phase noise characterization, the optical signals are converted to electrical signals by a high-speed photodetector with a 40-GHz bandwidth and the resulting repetition-frequency beat signals are analyzed with an electrical spectrum analyzer (R&S FSUP). The RF source signal is also characterized as a baseline reference.

## Data availability

All data generated or analyzed during this study are available with the paper. Source data are available from the corresponding authors upon reasonable request.

## Code availability

The code and algorithms used in this study are available from the corresponding authors upon reasonable request.

## Acknowledgements

We thank Liobate Technologies Ltd. for providing the TFLN electro-optic frequency comb chips and technical assistance. W. X. thanks Professor John E. Bowers of the Department of Electrical and Computer Engineering at the University of California, Santa Barbara, for valuable suggestions and discussions. The authors thank the Center for Advanced Electronic Materials and Devices (AEMD) of Shanghai Jiao Tong University for support in device fabrication. This work was supported by the National Natural Science Foundation of China (62175149, 62225407) and the National Science Fund Program for Distinguished Young Scholars (Overseas). The work was also sponsored by Huawei Technologies Co., Ltd.

## Author contributions

W. X. and B. Z. conceived the idea. The AlGaAs NOLM devices were designed by Z. L. and W. X., and fabricated by W. X. The numerical simulation and experiment were performed by Z. L. with assistance from Z. X., X. Z., and W. X. The results were analyzed by Z. L. and W. X. with fruitful discussions from Z. X., X. Z., and B. Z. The draft manuscript was written by Z. L. and W. X., and all authors contributed to the revision of the manuscript. The project was supervised by W. X., Y. S., and B. Z.

## Competing interests

The authors declare no competing interests.